\documentclass[12pt]{article}

\usepackage[english]{babel}

\usepackage{url}

\usepackage[a4paper,top=2cm,bottom=2cm,left=3cm,right=3cm,marginparwidth=1.75cm]{geometry}

\usepackage{amsmath}
\usepackage{graphicx}
\usepackage[colorlinks=true, allcolors=blue]{hyperref}

\usepackage{mathpazo}
\usepackage{microtype}

\title{Never eat a Pigeon with a Pumpkin: a model for the emergence and fixation of unsupported beliefs}
\author{Anders Sandberg, Institute for Futures Studies\footnote{anders.sandberg@iffs.se}\\ Len Fisher, University of Bristol}

\begin{document}
\maketitle

\begin{abstract}\noindent A popular poster from Myanmar lists food pairings that should be avoided, sometimes at all costs. Coconut and honey taken together, for example, are believed to cause nausea, while pork and curdled milk will induce diarrhea. Worst of all, according to the poster, many seemingly innocuous combinations that include jelly and coffee, beef and star fruit, or pigeon and pumpkin, are likely to kill the unwary consumer. But why are these innocuous combinations considered dangerous, even fatal? The answer is relevant, not just to food beliefs, but to social beliefs of many kinds. Here we describe the prevalence of food combination superstitions, and an opinion formation model simulating their emergence and fixation. We find that such food norms are influenced, not just by actual risks, but also by strong forces of cultural learning that can drive and lock in arbitrary rules, even in the face of contrary evidence.
\end{abstract}

\vfill 
\noindent This work was presented at the Oxford Symposium on Food and Cookery 2023, and a shortened form of the paper is Chapter 30 in {\em Food Rules and Rituals: Proceedings of the Oxford Symposium on Food and Cookery 2023}, ed. Mark McWilliams, Equinox eBooks Publishing, United Kingdom. pp. 294-306.
\vfill 

\section{Introduction}

There are foods that seem to go together, others that may not, as this poster from Myanmar dramatically claims:
\begin{center}
    \includegraphics[width=0.75\textwidth, angle=-90]{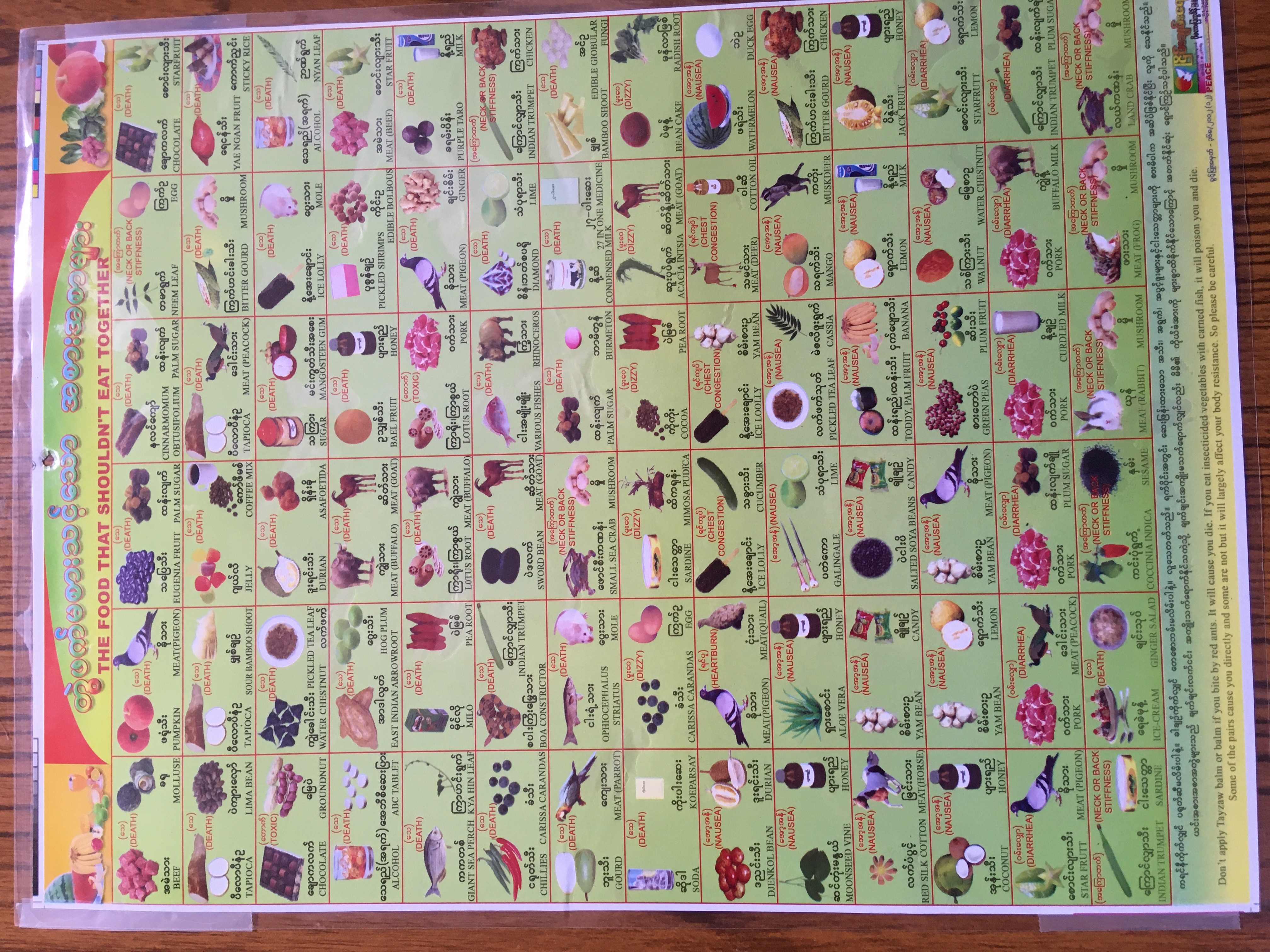} 
\end{center}

According to the beliefs represented here, pigeons and pumpkins are a definite no-no:
\begin{center}
     \includegraphics[width=0.5\textwidth]{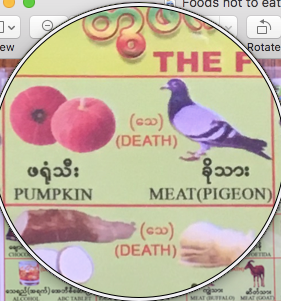}
\end{center}
As are fish and rhinoceros, not to mention iced lollies and moles, all combinations of which will result in instant death.

The poster is on display in many domestic kitchens and the beliefs that it represents are taken seriously. However, most of the food combinations claimed to be harmful are in fact innocuous \cite{van_hare_2008}. So, how can beliefs about their danger have developed and why do these beliefs remain current? Answers to these questions are likely to be relevant to wider questions of how social beliefs and cultural norms of many kinds can develop and become embedded, despite lack of evidence, or even contrary evidence, about their validity. Here, we apply basic statistical arguments and computer modelling of known cultural behaviours to help lay down the foundations for a possible answer.


\section{Background}
A cuisine is a way of structuring what is food or not, what is safe to eat or not, and what is appropriate to use or not. Within these bounds, no culture uses all edible food in its environment to make up its cuisine. Sometimes this is for good reason: some edible foods may be likely to be contaminated by pathogens, leading to food poisoning, while others may need treatment to remove toxins before becoming edible. Deep down, this is linked to "the omnivore's paradox" -- we need to subsist on many kinds of food and hence need diversity, yet each new ingredient is a potential risk \cite{fischler1988food}.

However, these practical reasons do not entirely account for the prevalence of food taboos. The British biochemist and virologist N.W. Pirie noted: "An interesting feature of such taboos is their persistence for centuries in communities which know that the tabooed food is eaten without harm by others."
\cite{pirie1984fluctuating}   

There are many possible reasons for the existence and persistence of food taboos \cite{meyer2009food} (see also \cite{navarrete2003meat}):
\begin{itemize}
    \item Food taboos for certain members of the society and to highlight special events
    \item Food taboos during pregnancy and food changes over the course of the menstrual cycle
    \item Food taboos as an ecological necessity to protect the resource
    \item Food taboos in order to monopolize a resource
    \item Food taboos as an expression of empathy
    \item Food taboos as a factor in group-cohesion and group-identity
\end{itemize}

We can divide this list into objective reasons (health, ecology), subjective reasons (religious, empathy), and social reasons (group cohesion and identity). Importantly, such rationalizations do not need to be correct in order to be effective. An ingredient that is generally believed to be dangerous may in fact be quite safe. A food taboo designed to increase group cohesion can become a source of conflict within and between groups. Preservation of a resource may have unforeseen knock-on effects on other resources. More generally, the overall world view of a culture may lead to mistaken conclusions about any or all of the factors listed above.

We may refer to such mistaken conclusions as "superstitions." A superstition in this context is not necessarily about supernatural things, but can be defined as an erroneous belief (i.e. does not conform with the actual facts) that causes some behaviour, often ritualistically \cite{vyse2020superstition}. Often this is defined as not conforming with the official or best known model of the world. Here, we will use superstition to refer to warnings about food that empirically do not have a commensurate actual risk.

One way in which such superstitions may become embedded is through the emphasis on the role of danger. As noted by Mary Douglas \cite{douglas2003purity}:
\begin{quotation}
\noindent It is obvious that a person when he finds his own convictions at variance with those of friends, either wavers or tries to convince the friends of their error. Attributing danger is one way of putting a subject above dispute. It also helps enforce conformity.
\end{quotation}
We will return to this by building an opinion model of food superstitions. 

Food superstitions may also originate as a form of costly, and hence hard-to-fake signalling of group loyalty or other desirable behaviour unrelated to the food. An interesting deliberate example was the rock band Van Halen's contract with concert venues that contained a ban on brown M\&{}M candy in the back stage area. While inconsequential, it was claimed to be a reliable test that the other complex and safety-relevant instructions were likely to be followed. \cite{roth1997crazy}\footnote{See also \cite{Mack2023why} for a critical view of this story.}
It is not hard to imagine that the rule could be misinterpreted as a taboo or a superstition by others (indeed, this is required for the M\&{}M trick to work), and then propagated to fans wanting to imitate. The actual origin of a superstition may be very different from how it functions later. 


This is not to say that superstitions are always irrational. If it is hard to tell safe from unsafe mushrooms and the cost of eating the wrong one is high, claiming all mushrooms are dangerous is a practical way of avoiding risk. Ascribing the ill effect of eating certain mushrooms with alcohol to an erroneous theory of digestion still produces the correct behaviour. It is common to hear claims that food superstitions are based on a factual basis \cite{douglas2003purity}, but many clearly do not and do cause irrational behaviour, sometimes with detrimental nutrition and health effects \cite{ekwochi2016food}. Avoiding boiled water due to it being a "hot" food when treating people with fever causes avoidable harm. 

Another cause may be cultural copying. Human survival is dependent on many fairly opaque practices, and copying behaviours in detail rather than experimenting is in many cases the smart choice \cite{boyd2011cultural,henrich2015secret}. Hence random superstitions persist because it is often counterproductive to investigate them.

\subsection{The Myanmar poster}

The Myanmar poster provides a particularly dramatic example of a set of food superstitions that have emerged and apparently remained stable over time. Its origins are unclear, although the beliefs represented extend back for at least eighty years, and perhaps much longer\footnote{\url{https://www.vice.com/en/article/gyw9eq/myanmar-food-superstitions}}.

Burmese medicine divides food into heating or cooling, similar to the Chinese classification in food medicine. This makes food matching and balancing a natural idea. 

Myanmar is also at a cultural and developmental cross-roads, with influences from Arabic, Chinese and Indian traditions. In its present parlous condition, considerations of safety are paramount, and this may perhaps extend to food safety. 
\cite{downs2019interface}
It is thus natural to ask whether the emphasis on and choice of "unsafe" food pairings in the Myanmar poster may have emerged from the influence and/or beliefs of some other culture. We began our consideration of the beliefs exemplified in Myanmar poster by examining the food pairing beliefs of other cultures, and asking whether any of these beliefs may have led to those on which the Myanmar poster is based.

\subsubsection{Risks of the food on the poster}

Several of the ingredients are potentially risky to eat. Most obviously, some are medications or herbs used for medicine. For example, the whole moonseed wine plant is toxic, and starfruit contains oxalic acid. There are also ingredients that must be prepared correctly. Yam beans have edible roots, but the rest of the plant is toxic. Some ingredients could potentially go bad, most obviously meat and seafood. Peanuts in a tropical climate may be a source of aflatoxins. However, as we will see, relatively few if any ingredients are known or likely to be risky as combinations. 

\subsubsection{Structure of the poster pairings}

\begin{figure}[h]
\begin{center}
    \includegraphics[width=\textwidth]{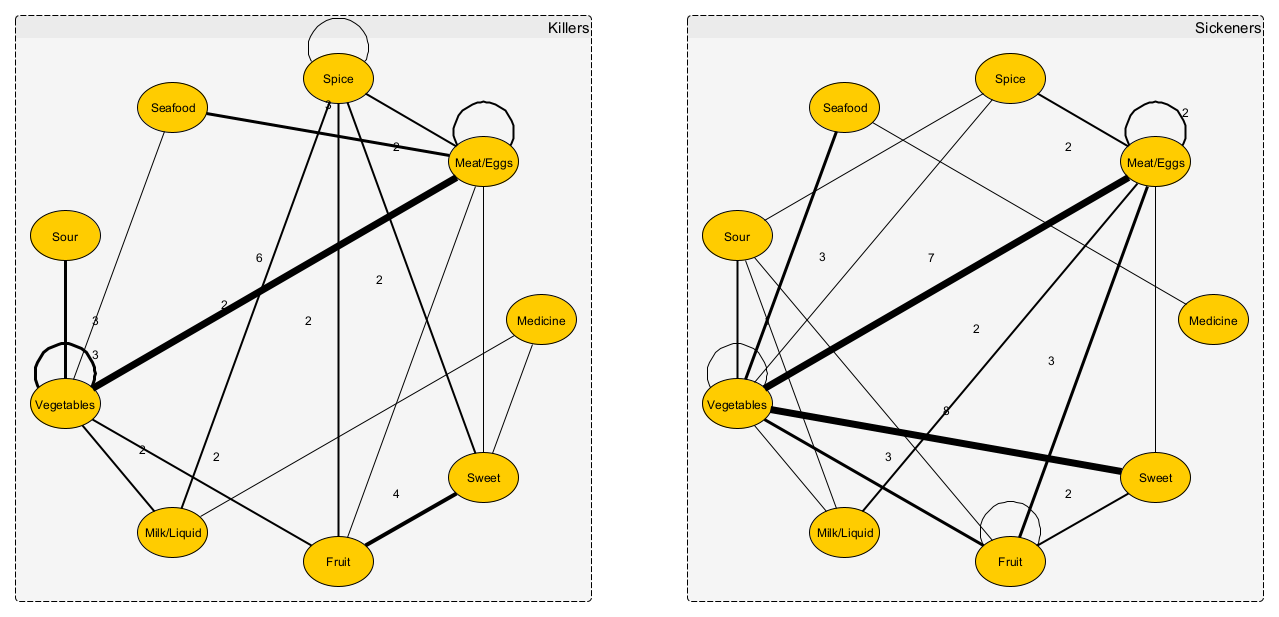} 
\end{center}
    \caption{Graph of food interactions in the poster. Each line corresponds to a set of warnings, with width proportional to the number of cases. The left-hand graph represents claimed deadly interactions, the right-hand graph merely sickening ones.}
    \label{fig:graph}
\end{figure}

Is there any particular structure in the pairings the poster warns against? We can order the ingredients after rough type and separate the deadly and merely sickening/toxic combinations, producing an interaction graph (figure  \ref{fig:graph}).

There are some obvious patterns. Vegetables and meat combinations are strongly represented in both cases. However, vegetable-sweet combinations are well represented among sickeners but not killers. The other category pairings have a different pattern but given the small sample it is risky to draw too strong conclusions.\footnote{Different categories have different frequency, so merely counting co-occurrences can be misleading. A more principled way of analysing the weights of the graph is the mutual information components between them. The mutual information component between category $i$ and $j$ in these graphs $I(i,j)=\log(\Pr[i,j]/\Pr[i]\Pr[j])$ indicates how much one can predict that category $i$ is the other category of food if told that category $j$ is in a dangerous pairing (the full mutual information is the sum of these weighted by $\Pr[i,j]$). In this case the maximal mutual information components corresponds to the strongest links: they remain salient even when taking into account the prevalences of the categories.}

There is no clear pattern in what meats are killers or sickeners. 
Ayrvedic incopmpatibilities like fruit with milk are not found on this list

\subsection{Taboo pairings}

While taboos against particular ingredients are widespread, combination taboos are more rare. However, they are still found in many places. 

\subsubsection{Religious taboos}

Jewish kashrut rules against eating meat with dairy, or even using utensils that have been used for preparing or eating the other kind of food. There is also a required wait between eating meat and dairy for a period from three of six hours. While there have been attempts by e.g. Maimonides to give a rational health justification  such justifications easily fall apart. 
Other suggested motivations have been  Judaism's desire to separate categories of life from death \cite{douglas2003purity}, or the importance of the food rules as a way of preserving cultural identity and difference. 

It is worth noting that in some places Jewish communities intermingled with local communities and may have left the meat-milk taboo in place even when the community composition changed.

\subsubsection{Theory-based combination taboos}


Ayrvedic theories about digestion produce a plethora of food combination warnings (as well as warnings about incompatible places, time and other categories). Food is divided into a number of categories that are digested in different orders and at different speed, and this process can be derailed by eating the wrong combinations.\footnote{One intriguing finding is that Indian cuisine is characterized by a strong negative flavor pairing pattern: pairs of ingredients in a recipe are less likely if they have overlapping flavor profiles, unlike in most other cuisines. This may have been due to medicinal beliefs \cite{jain2015spices,jain2015analysis}.}
Melons are risky with any other food. Grains and fruit, beans and meat, cheese and eggs, and many other pairs cannot be combined. Milk and fish is claimed to cause vitiligo.
Viruddha Ahara (wrong food combination) is regarded as a source of illnesses, and in modern academic Ayrvedic medicine motivated by particular biochemical changes \cite{ashok2022viruddha} (where the existence of a few cases are used to justify the entire traditional system without further evidence).
\footnote{\url{https://theprint.in/opinion/food-combining-ayurvedic-taboos-are-back-as-fad-its-ok-to-mix-fruits-with-vegetables/772070/}}

These rules have made their way into modern alternative medicine and diet fashions via mid-1800s "trophology". 

As an example, some versions of the Paleo diet also involve food pairing rules like avoiding consuming carbs/starches with protein, don't mix different types of protein, don't pair acidic foods with starches, don't eat fruit and vegetables at the same time\footnote{\url{https://ultimatepaleoguide.com/food-combining/}}. This is motivated by the beliefs that foods digest at different speeds, need different enzymes to break them down and this is affected by food-affected pH, and that food may ferment in the stomach. There is no scientific support for these beliefs, and they seem to come from a mix of traditions and common-sense reasoning about how digestion might work\footnote{\url{https://www.healthline.com/nutrition/food-combining}}.

Combining seafood and cheese is avoided in some places, apparently a regional Italian view that got exported post-WW II, possibly even originating with Aristotle's view that cheese and fish are digested at different speeds\footnote{\url{https://www.atlasobscura.com/articles/mixing-seafood-and-cheese} }.


In China the "conflict food" theory argues that eggs and soya milk, honey and Chinese onions, and other combinations are harmful\footnote{\url{http://en.people.cn/90782/8436029.html}}. This is rooted in traditional Chinese medicine, where different foods are assigned to hot/yang and cool/yin, different phases of the cycle of ki, and a balanced meal combines them (while, when used as medicine, the meal should compensate for the imbalance of the patient). These ideas are widespread outside China in Asia, although the assignment of ingredients to categories varies. 

In Japan this led to lists of "[things] to be eaten [or not to be eaten] together" (kuiawase) in texts from the Edo period and later that are similar in form, if not content, to the Myanmar poster \cite{kinski2009admonitions,schlachet2018nourishing}. Many of these combinations also appear based on intuitive ideas like that oil and water don't mix, so tempura (oily) and watermelon (watery) should not be combined\footnote{\url{https://japantoday.com/category/features/food/traditional-bad-food-combos-in-japan}}. These lists were however also shaped on ideas about economics and social stability rather than just bodily health.

\subsubsection{Fruit and milk}

Fruit and milk combination taboos are common worldwide. 

In Iran eating watermelon after milk is seen as risking stomachache\footnote{\url{https://skeptics.stackexchange.com/questions/50725/is-eating-watermelon-after-milk-bad}}, perhaps a belief imported from India. 

Mango and milk is seen as risky in Brazil.
This is sometimes extended to other fruit, eggs, or meats with milk \cite{trigo1989food,erickson1990hungry}.
One widely repeated modern explanation is to get slaves (often eating mango) to avoid drinking expensive milk!\footnote{\url{https://www.lsg-group.com/news/global-food-myths-3-are-mango-and-milk-harmful-together/}}

In the US cherries with milk are sometimes seen as harmful. One possible cause may have been the death of president Zachary Taylor in 1850 after eating cherries and drinking iced milk.
See also \cite{parenti1998strange} for the argument that it may have been poisoning (!) and that there is no evidence for the cherries and milk in historical sources. Regardless of what actually happened and what he actually ate, the milk and cherries story appear to have become widely publicized, creating the idea that the combination is risky. 

Another (possibly originally German) belief is that cherries and water is risky. Modern versions invoke that the water dilutes digestive enzymes or stomach acid that kills bacteria \cite{gigerenzer2015calculated} (but why would it then be just cherries that are risky?)

There are Malay warnings against combining Chinese melon with honey, and papaya with tapai (a local sour dough snack) \cite{mckay1971food}.

In 2013 a 3-year old Chinese girl died shortly after consuming persimmons and milk. While the role of the milk was equivocal,  this appears to have started rumors that persimmons are risky together with milk, crab or other ingredients\footnote{\url{https://skeptics.stackexchange.com/questions/52792/is-drinking-milk-after-ingesting-persimmon-dangerous}}. Unripe persimmons (like some other food) do contain tannins that can cause hard bezoar stones to form in the stomach \cite{iwamuro2015review}.

\subsubsection{Dangerous drink}

In Belgium and the Netherlands there is a belief that one should not mix tonic and Baileys Irish Cream, since it would form stones in the stomach. Here the belief likely originates from the fact that the acidic tonic causes the casein in the cream liqueur  to coagulate, and the visible curdling motivates the belief. 
This is also a plausible reason for claims milk cannot be combined with fruit or other acidic food in many cultures.

Generally milk is regarded as risky to combine with most ingredients in the Ayrvedic tradition, as well as in Spain ("Después de la leche, nada eches"). Given the prevalence of lactose intolerance this may not be surprising, but note that the belief is about the combination of milk with other ingredients rather than the milk itself.

There are also claims that mixing different types of alcohol makes hangover worse ("do not mix the grain with the grape"). This could have emerged simply because drinking {\em more} makes hangovers worse. There are also beliefs that the order matter: "Beer before liquor, never been sicker; liquor before beer, you’re in the clear"\footnote{\url{https://www.healthline.com/nutrition/beer-before-liquor}}. Again, the amount matters more. 

In Switzerland there is a belief that drinking cold drinks (except alcohol) with cheese fondue is dangerous, again  motivated by concern about molten cheese forming lumps \cite{heinrich2010effect}.

\subsubsection{Actually risky pairings}

Given this plethora of false beliefs, are there any food combinations that are actually risky? They certainly exist, although they may be rarer and usually weaker than one would expect given the chemical complexity of food \cite{van2021toxicology}.

Mushrooms like morels and ink caps are a well-known bad combination with alcohol in Western cooking \cite{groves1964poisoning,michelot1992poisoning}. At least the ink cap contains coprine that  inhibits aldehyde dehydrogenase, producing an antabus-like effect (coprinus syndrome) lasting 24 to 72 hours with flushing, nausea, vomiting and rapid heartbeat 15 min after ethanol ingestion. Coprine or similar compounds are found in other species \cite{haberl2011case}, producing the general advice to avoid alcohol with wild mushrooms.

In Singapore and other southeast Asian countries the combination of durian fruit and alcohol are assumed to be dangerous \cite{mckay1971food}, and deaths have been ascribed to it\footnote{\url{https://www.malaymail.com/news/life/2019/08/16/thai-man-found-dead-at-bus-stop-allegedly-after-eating-durian-with-wine/1781362}}. A study suggested that sulphur compounds like diethyl disulfide in the fruit inhibit alcohol breakdown, giving some rationale for the warning \cite{maninang2009inhibition}. Other kinds of fruit such as lemons, {\em Averrhoa carambola}, and {\em Syzygium samarangense} also appear to inhibit part of the alcohol breakdown chain \cite{zhang2016effects}, while other fruit may help it and reduce hangover \cite{srinivasan2019influence}. However, the effects on the enzymes and alcohol breakdown in these studies are relatively modest.\footnote{There is also a Malay belief that durian cannot be safely combined with cassava and chloroquine. As far as we know there has not been scientific investigation of these interactions. \cite{mckay1971food} }

We have heard claims that eating cassava/manioc with ginger is risky, since ginger enzymes may release cyanide from the linamarin. However, since this happen by mere hydrolysis without ginger, the cassava needs to be treated anyway: the ginger appears to be irrelevant unless the cassava is badly prepared. 

There have been concern that eating fish with leafy vegetables like spinach could lead to the formation of nitrosamines. Nitrosamines form in the stomach from ingesting nitrites that are found in various ingredients. Processed meat and smoked fish (sources of nitrite) are often combined with salt that can damage stomach mucosa and contribute to carcinogenesis. However, the evidence is fairly limited \cite{jakszyn2006nitrosamine}.

These examples mostly do not have the same cultural legs as the previous section. It is also notable that these pairings are far fewer than the imputed risky pairings believed above.

Beside the combinations, there are far more ingredients that in themselves have risky properties. Persimmon and alcohol have already been mentioned. Ingredients may have a high risk of food-borne illness (e.g. poultry, fish, shellfish, raw vegetables), be toxic when cooked the wrong way (e.g. cassava, kidney beans, ackee fruit, pangi fruit), have pharmacological side effects (e.g. various herbs, grapefruit inhibiting drug breakdown, activated charcoal in black drinks blocking medications, tyramine in wine and cheese causing dangerous blood pressure spikes in people taking MAO-inhibitors), reduce nutrient uptake (e.g. phytic acid in legumes decreasing metal absorption) and so on. Some ingredients may also be risky for some people due to allergies or other toxicities (e.g. oxalic acid in star fruit causing kidney damage in sensitive people \cite{lee2012star}), not to mention contamination (e.g. pesticides or adulterants).

The World Health Organization estimates that 600 million (about 1 in 10) people in the world fall ill after eating contaminated food and 420,000 die every year\footnote{\url{https://www.who.int/news-room/fact-sheets/detail/food-safety}}. This doesn't include the effects of unhealthy eating in general on long-term health.
Hence there are reasons to be wary of many ingredients and at least some combinations.
There are naturally also many food pairings that are shunned simply because of bad taste. That poor combination may then become entrenched as a rule of "do not combine X and Y" without explanation, morphing into a belief of actual risk.\footnote{One future example may be Italian prescriptions that dishes with garlic should use pecorino cheese, never Parmesan cheese (usually motivated by considerations of taste). This might plausibly morph into a taboo similar to avoiding grated cheese on seafood, that starts as a matter of taste but gradually becomes seen as a health issue as a digestion "explanation" is added.}


\subsubsection{Summary}

Food pairing taboos/superstitions occur in many cultures. Some general observations from the above examples are: 
\begin{itemize}
\item Most of the banned combinations are motivated by health reasons rather than religious ideas. 
\item Many of these are explained by a pre-scientific theory of how digestion or nutrition works.
\item Some of these are likely inspired or strengthened by observable external interactions such as milk curdling in acidic environments or molten cheese hardening in contact with liquid.
\item Cases of death by food poisoning are noticeable and can cause taboo pairings to spring up. 
\end{itemize}

\subsubsection{Pre-Conclusion}

We can find no convincing evidence for real danger (certainly not danger of death) in \textit{any} of the pairings listed (with the exception of ingredients like moonseed vine that can be dangerous on their own). Nor, despite a careful comparison, can we find any convincing relationship between the claimed "dangerous" food pairings in the Myanmar poster and the food pairing beliefs of the Ayurvedic, traditional Chinese or Japanese, or any other belief system concerned with food pairings. It seems that we must look elsewhere for the origins and cultural embedding of these beliefs.

\section{Models}

When it comes to a fresh search for potential explanations, we are the beginning of a journey. What we offer here is very preliminary, but promising.
We have opted for two paths. The first is to examine the statistics of the food pairings in terms of categories into which the individual foods may most readily be placed (meat and eggs, vegetables, seafood, fruits, spices, milk and milky foods, sweeteners, contributors of sourness, and medicines). Preliminary results are encouraging, especially in revealing differences between "killer" foods and those believed only to cause some form of illness. 

Our second path is to develop a simple statistically based model of how beliefs may originate, spread and then become fixated in a culture even when there is little or no evidence for their validity. This path is wider and more important than may appear at first sight, since similar processes may be involved in the development of all sorts of belief -- political, social and religious beliefs being among them.

There are many ways in which relatively unsupported beliefs may become embedded. One strong candidate is the illusion of control effect: people often think they have more control over a situation  than they actually have. Accidental correlations are taken as signs of causal impact, and learned as ways of control a situation. In this case the actual risk of food is largely random, and the decision to eat it has little effect on personal risk. However, imagined patterns are easily found and can then be transmitted and reinforced socially. 


We will start with a sub-model of how observing food poisoning may originate a belief among people that a certain food pairing is risky. Then we will present a sub-model of how beliefs spread in a social network. After that we will combine them into a joint model of how food taboos and other unsupported beliefs may emerge, spread and become embedded in a culture. 

\subsection{Observational evidence}

Actual observations of people falling ill after eating an ingredient is evidence that it is unsafe. If we assume there is a case of food poisoning the people hearing about it will update towards the ingredient being unsafe; if they think it is unsafe and they hear somebody survived eating it, they may update in the opposite direction. 

The ideally rational person updates beliefs according to Bayes' rule. This is a vast oversimplification, but can show how even such persons may become concerned about a food combination.

Consider the probability $p_i$ of falling ill by eating ingredient $i$. At first this might be unknown, making people start with a uniform prior for the actual value of $p_i$. After observing $n$ people eating it with $k$ of them falling sick, the posterior probability distribution of $p_i$ will be \begin{equation}
    f(x) = \frac{x^{n-k}(1-x)^{k}}{B(n-k+1,k+1)}
\end{equation}
where $B(\alpha,\beta)$ is the incomplete gamma function. This gives a mean estimate \begin{equation}
    \langle p_i \rangle=\frac{n-k+1}{n+2}.
\end{equation}

If we go from $k=0$ to $k=1$ (we hear about a case of somebody falling ill eating a particular ingredient) the change in estimate is 
$1/(n+2)$: if there is a lot of evidence that the ingredient is safe ($n$ large), the update is small. Maybe there is some risk, but it must be small since otherwise it contradicts experience. If $n$ is smaller, then the update becomes much larger and more salient.

This may explain why food combinations may appear more risky than basic ingredients: many have eaten pigeon or pumpkin, so they are known to be fairly safe. But few have encountered the combination, so if its safety is treated as separate from the safety of the ingredients any story about a fatal meal produces a strong update of belief. 

Consider the case of having seen 1000 instances of eating pigeon, and 1000 instances of eating pumpkin. If there are 50 possible choices of ingredients and they co-occur randomly there will be 20 cases of pigeon plus pumpkin.

Over a year a person eats on the order of 1000 meals. If the annual risk of food poisoning is 10\%, that makes the risk per meal $p\approx 0.0001$. So there is just 10\% risk for each ingredient to have one or more cases of illness in the above case. Hence if there is a single case of illness when eating pigeon and it happens to have been with pumpkin, the update of the safety of pigeon and pumpkin individually is from 1000/1001 (999 safe cases) to 1000/1002 (999 safe, 1 unsafe), a change in probability close to 1 in 1000. But for the {\em combination} it is from 20/21 to 20/22, a change of 4 in 100, 43 times larger!

A statistically careful person will recognize that the variance of the pigeon-pumpkin estimate is higher than for the single ingredients and hence one should be more cautious in acting on it, due to its high uncertainty. 
However, humans are not ideal rational agents nor statistically careful.

This update is likely amplified by known cognitive biases. Something being riskier than expected is highly salient. The conjunction fallacy makes it easy to overestimate the probability of X and Y compared to X and Y on their own ($\Pr[X \cup Y]\leq \Pr[X]$ and $\leq \Pr[Y]$). The availability heuristic makes easy to recall events seem more likely, and later confirmation bias makes supporting evidence appear more salient and relevant than the disconfirming evidence. The representativeness heuristic amplifies the base rate fallacy, making people overestimate the risk from the combination. Generally we prefer certainty (the zero-risk bias), making avoiding the potentially risky combination look reasonable. In addition, the local food poisoning tragedy will be experienced socially not just by one person but by several, all reacting roughly in a similar way. The end result can be a nucleus of risk estimates overestimating the food combination risk that then convinces the rest of the community by bandwagon effects.


\subsection{Lifelong opinion}

The simplest possible opinion model has each person retain their views throughout life, being "born" with a view that is a random mixture of parents and family members. This model has no actual input from reality about actual food risk and each ingredient is treated entirely separately. Adding opinions about combinations of ingredients however works exactly the same.

Person $X_i$ ($i=1,\ldots N$) believes the safety of food $j$ is $S_{ij}\in \{0,1\}$ (unsafe/safe).  The probability that they will think $S_{ij}=1$ is based on the average view of who they know \begin{equation}
    \Pr[S_{ij}=1]=\langle S_{kj}\rangle_k
\end{equation} where $k$ ranges over their social circle (which can be the entire population). We can model the opinion change by replacing a random old person with a new-born person gaining opinions randomly by the above formula at each step. 

\begin{figure}[h]
\begin{center}
    \includegraphics[width=\textwidth]{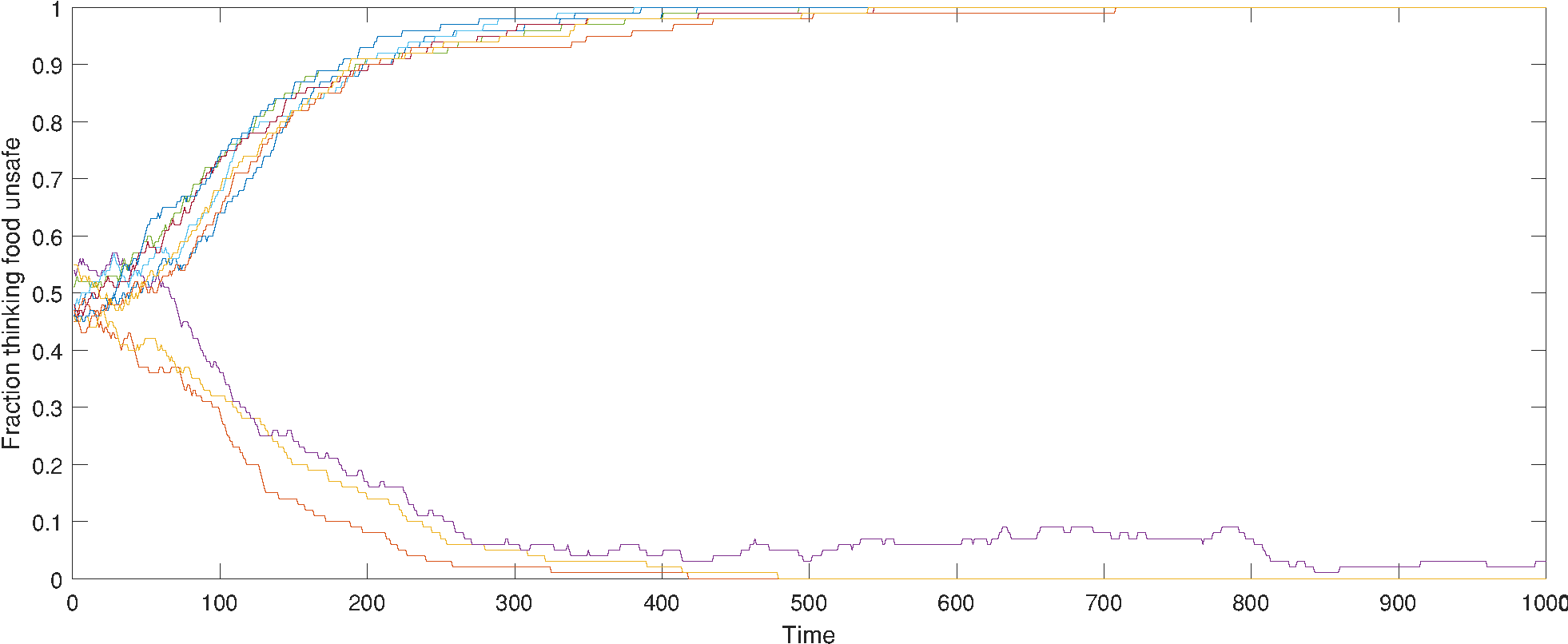} 
\end{center}
    \caption{10 realizations of the lifelong opinion model. Starting with an even mix of safety views, random copying of views between members of a society leads fixation where everybody agrees it is safe or unsafe.}
\label{fig:lifelong}
\end{figure}

The end result is that the initial spread of opinions on each food will gradually narrow, with each food eventually seen as entirely safe or unsafe by everyone (figure \ref{fig:lifelong}). Which outcome happens is random, although biased by the initial composition of the views. 

This is basically a genetic fixation model of an allele under neutral selection in a monoploid population, a well studied topic in population genetics. Drawing from results in that theory, the expected generation time to coalescence is $N$. Since a generation is a full replacement of the population it will take $N^2$ individual replacement steps to reach fixation. In practice the time to fixation will have an exponential probability distribution with this mean and a wide variance of $N^4$.
The probability that a lone dissenter will eventually convince the entire society about their view for an ingredient is $1/N$.


Adding noise, where there is a certain "mutation rate" where people change their minds, leads to eternal drift although the effect is (for a small rate) still near-convergence. Consensus never becomes perfect, and rarely views on an ingredient flip. This is also equivalent to the situation if there is an inflow of newcomers with different beliefs. For a high rate views are driven by the noise, performing a random walk at intermediate levels of belief.

If the population size increases the time to fixation grows, while shrinking populations more rapidly converge \cite{waxman2012population}. Societies that have undergone a low $N$ population bottleneck (e.g. founding settlers, a past disaster) are hence more likely to have fixated beliefs. This is also true for minorities.

Myanmar, with its many local ethnicities with presumably small $N$, would appear to be a good candidate for such a process generating fixation.

\subsection{Observation and copying}

We can now combine the observational effect, where people try to estimate the riskiness of food (and stay away from risky combinations) and also copy the attitudes of others in their social milieu. 

There are $N$ people. Each has a risk estimate for an ingredient or combination, $0\leq R_{ij}\leq 1 $, where $i$ is the person and $j$ the food. 

There is a fixed probability $P_{illness}=1/1000$ that a person will fall ill after eating, independent of the food (there are no actually safe or unsafe foods in this model).

Each day each person eats a dish $j$, which is randomly selected. With probability $\min(1,P_{pickyness}R_{ij})$ they select another dish (repeating until they find one they think are safe). $P_{pickyness}\geq 0$ represent how cautious people are about risky food.

If they fall ill they update their risk estimate as \begin{equation}
    R_{ij}\leftarrow (1-c_{evidenceUpdateBad})R_{ij} + c_{evidenceUpdateBad}
\end{equation} where $0\geq c_{evidenceUpdateBad}\geq 1$ represents how strongly beliefs are updated by a bad experience. If they do not fall ill the update is instead update as \begin{equation}
R_{ij}\leftarrow (1-c_{evidenceUpdateOK})R_{ij} + c_{evidenceUpdateOK}\end{equation} where $0\geq c_{evidenceUpdateOK}\geq 1$ represents how strongly beliefs are updated by an OK experience. Plausibly $ c_{evidenceUpdateOK}< c_{evidenceUpdateBad}$ due to saliency.

Finally, each day each person tells $i$ somebody else $k$ (randomly selected) about a random food $j$, causing an update that depends on whether it is a warning about something being riskier than they think or whether it is safer than they think:
\begin{equation}
    R_{kj}\leftarrow 
    \begin{cases}
   (1-c_{opinionUpdateOK})R_{kj} + c_{opinionUpdateOK} R_{ij},& \text{if } R_{ij}\leq R_{kj}\\
   (1-c_{opinionUpdateBad})R_{kj} + c_{opinionUpdateBad} R_{ij},              & R_{ij}>R_{kj} \, \text{otherwise}
\end{cases}
\end{equation}

This is an opinion model where there can be ratchet effects both from avoiding perceived risky food (reducing the chance of disconfirming evidence in the future), from the salience of bad outcomes, and preferential spread of warnings rather than reassurances. 

If $c_{opinionUpdateBad}=c_{opinionUpdateOK}=1$ we get a version of the previous model where opinions are merely copied. 

\begin{figure}[h]
\begin{center}
    \includegraphics[width=\textwidth]{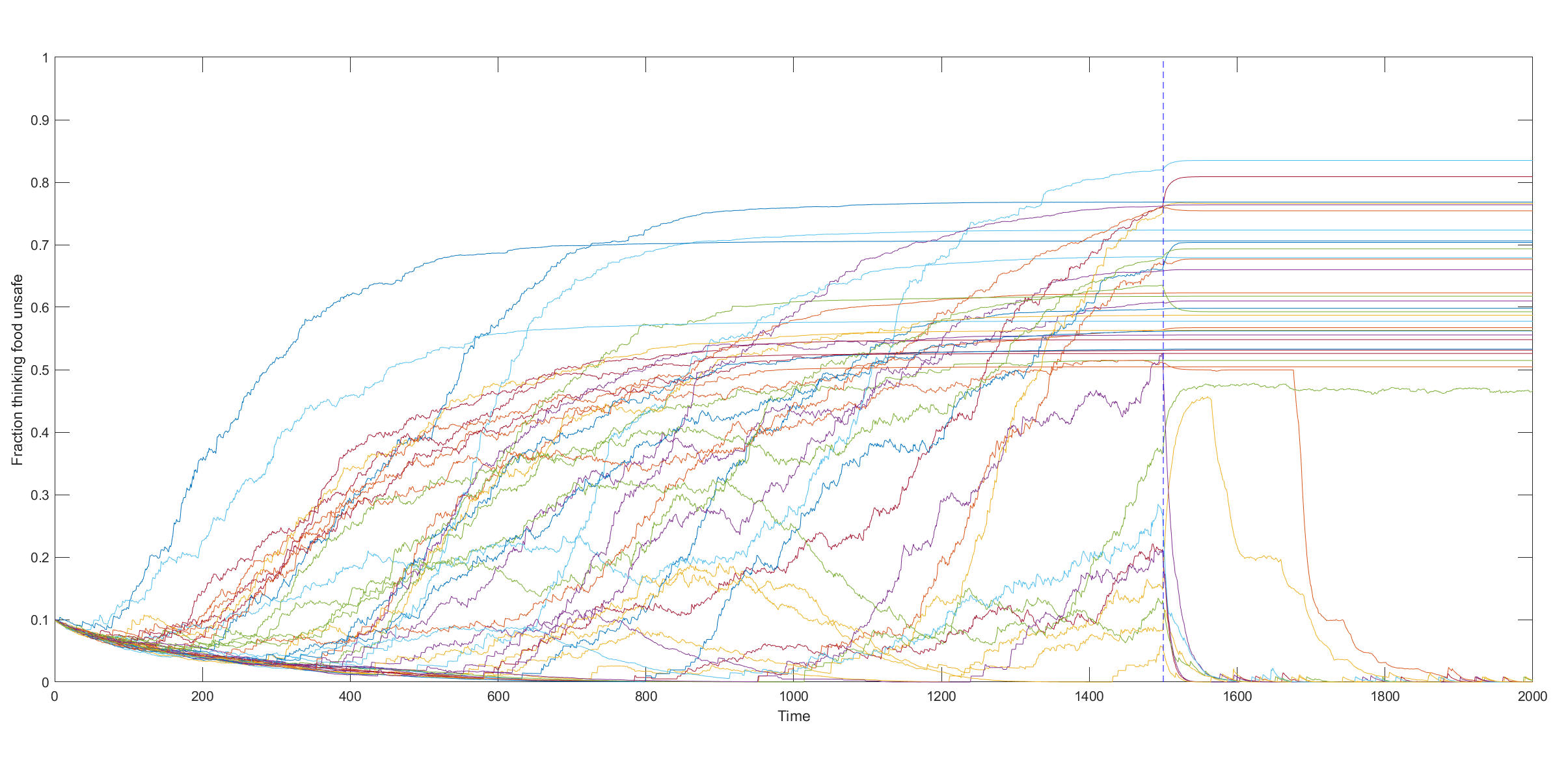} 
\end{center}
    \caption{Full model of food risk estimation using observation and copying. The vertical axis shows the fraction of the population believing a given combination (marked by color) is unsafe as a function of time. At time 1500 a broadcast of opinion similar to the poster is introduced.  $P_pickyness=2$, $c_{evidenceUpdateOK}=0.3$, $c_{evidenceUpdateBad}=0.95$, $c_{opinionUpdateOK}=0.3$, $c_{opinionUpdateBad}=0.95$, $c_{opinionUpdatePoster}=0.1$.  }
\label{fig:full}
\end{figure}

Running a simulation with 100 people and 40 ingredients produced the dynamics in figure \ref{fig:full}.  Opinions move around randomly as stories about people falling ill circulate or are refuted by other experience. Due to picky eating avoided food risk estimates remain fairly stable since counterexamples are rare. In this small simulation a slowly changing pattern of food preferences emerges.

\subsection{Broadcasting of opinion}

As a variation, we can also add perceived ‘authoritative’ broadcasting of views, as in the Myanmar poster. Let us assume that each day, in addition to the individual interactions, there is an opinion update from such an opinion originator (here arbitrarily selected as $i=1$)
\begin{equation}
    R_{kj}\leftarrow 
   (1-c_{opinionUpdatePoster})R_{kj} + c_{opinionUpdatePoster} R_{ij}
\end{equation}

In the above figure the poster appears at time 1500. The effect is incredibly striking.  Even for a very small value of $c_{opinionUpdatePoster}=0.1$ the risk estimates rapidly converges across the population into a largely fixed structure. While each individual may be far less swayed by a poster than the vivid story from a friend, the entire population is simultaneously pushed towards a particular view and this will also affect people in their one-to-one conversations. 

\section{Conclusions}




Our model offers, in principle, an explanation of how factually unsupported, or minimially supported beliefs may emerge, become embedded, and remain stable in a culture, often against contrary evidence. We do not claim that this is the only process involved. In the Myanmar case, for example, the country  has long suffered unrest and conflict. It is a psychological commonplace that observing superstitions often helps regulate tension in uncertain and stressful situations by creating a feeling of control. It is not implausible that everyday fear can be ameliorated by adhering to a clear set of kitchen rules, at least at home. This would strengthen the previous processes, perhaps helping explain why the rules are so singularly firm. 

The process described by our model does however represent what seems likely to be a more general and widespread process of how such a set of beliefs may emerge: in the example case, the natural human tendency to divide food into safe and risky, the salience of food combinations given the big intellectual frameworks of Indian and Chinese medicine, the random emergence of individual beliefs from experiences,  amplification by common cognitive biases and heuristics, the social imitation of others leading to shared beliefs, and finally encoding as a formal poster that further broadcasts and reinforces the pattern. Similar processes are also likely to be at work in the emergence, embedding and stabilization of political, religious, economic and other social beliefs.

\section*{Acknowledgments}
We have particularly valued the input, expertise and advice of Naomi Duguid and Priya Mani on Myanmar foods and food customs. We also wish to thank 
Paulo Abelha,
Peter Svensson,
Joshua Fox,
David Mannheim,
Lisa Enckell,
meismarv,
ketil,
A.V. Turchin-Bogemsky,
Victor Crespo,
and Vítězslav Ackermann Ferko
for input on Twitter about food pairing taboos in their own cultures. Further, we would like to thank the participants at the 2023 Oxford Food and Cookery Symposium for comments and support. 

\bibliographystyle{apalike}
\bibliography{sample}

\end{document}